\documentclass[journal]{IEEEtran}

 \usepackage{multirow}
 \usepackage{lscape}
\usepackage[english]{babel}

\usepackage[letterpaper,top=2cm,bottom=2cm,left=2cm,right=2cm,marginparwidth=1.75cm]{geometry}

\usepackage{subcaption}
\usepackage{amsmath}
\usepackage{graphicx}
\usepackage[colorlinks=true, allcolors=blue]{hyperref}

\usepackage{epstopdf}

\usepackage{varioref}
\usepackage{float}
\usepackage{comment}

\usepackage{xcolor}
\usepackage[linesnumbered,ruled,vlined]{algorithm2e}
\usepackage{amsmath}
\usepackage{caption}
\usepackage{graphicx}
\usepackage{hyperref}

\SetCommentSty{mycommfont}

\SetKwInput{KwInput}{Input}                
\SetKwInput{KwOutput}{Output}              

\usepackage{flushend}
\usepackage{tabularx,booktabs,textcomp}
\newcolumntype{C}{>{\centering\arraybackslash}X} 
\newcolumntype{L}{>{\raggedright\arraybackslash}X} 
\usepackage[noadjust]{cite}

\usepackage{wasysym}
\usepackage{lipsum} 
\usepackage{setspace}
\usepackage{newtxtext,newtxmath}

\usepackage{multirow}
\usepackage{rotating}
\title{Task Offloading for Smart Glasses in Healthcare: Enhancing Detection of Elevated Body Temperature}

\begin{document}

\author{Naouri Abdenacer, Nouri Nabil Abdelkader, Attia Qammar, Feifei Shi, Huansheng Ning, Senior Member, IEEE, and  Sahraoui Dhelim.
	\thanks{Naouri Abdenacer, Attia Qammar, Feifei Shi and Huansheng Ning are with the University of Science and Technology Beijing, Beijing 100083, China}
	\thanks{Nouri Nabil Abdelkader is with the University of Djelfa, Algeria}
    \thanks{Sahraoui Dhelim is with the School of Computer Science, University College Dublin, Ireland}
	\thanks{Corresponding author: Huansheng Ning (ninghuansheng@ustb.edu.cn).}
}

\maketitle

\begin{abstract}
Wearable devices like smart glasses have gained popularity across various applications. However, their limited computational capabilities pose challenges for tasks that require extensive processing, such as image and video processing, leading to drained device batteries. To address this, offloading such tasks to nearby powerful remote devices, such as mobile devices or remote servers, has emerged as a promising solution. This paper focuses on analyzing task-offloading scenarios for a healthcare monitoring application performed on smart wearable glasses, aiming to identify the optimal conditions for offloading. The study evaluates performance metrics including task completion time, computing capabilities, and energy consumption under realistic conditions. A specific use case is explored within an indoor area like an airport, where security agents wearing smart glasses to detect elevated body temperature in individuals, potentially indicating COVID-19. The findings highlight the potential benefits of task offloading for wearable devices in healthcare settings, demonstrating its practicality and relevance.

\end{abstract}

\section{Introduction}
Wearable devices, such as smart glasses like Google Glasses, have gained significant popularity in recent years, thanks to their wide range of applications and functionalities. These innovative devices have demonstrated their potential in various fields, including augmented reality (AR), healthcare, and security, offering users immersive experiences and enhanced capabilities. However, one of the major challenges faced by wearable devices is their limited computational capabilities. Despite their compact and portable nature, these devices often struggle to handle resource-intensive tasks, such as image and video processing. These tasks demand substantial computational resources, resulting in increased power consumption and rapid battery drain. As a consequence, the user experience is compromised, hindering the seamless execution of complex applications and services. To address these limitations and improve the overall performance of wearable devices, researchers have delved into the concept of task offloading. Task offloading presents a promising solution by allowing the transfer of computationally intensive tasks from wearable devices to nearby devices with higher processing capabilities. These remote devices can include mobile devices or edge servers, which possess greater computational power and resources.

By leveraging the computational capabilities of nearby devices through edge computing and fog computing, wearable devices can enhance the user experience and extend battery life. Task offloading allows wearables to transfer resource-intensive tasks to remote devices, reducing the strain on the wearable's limited resources. This redistribution of computational load results in improved performance, faster task execution, and prolonged battery longevity. With edge and fog computing, wearables can efficiently offload tasks, conserve energy, and deliver advanced applications with faster response times \cite{contess}.

Authors in \cite{piwek2016rise} have explored the promises and barriers associated with consumer health wearables, shedding light on the challenges related to security, privacy concerns, form factors, weight, and comfort. Authors in \cite{varghese2016challenges} have investigated in edge computing, highlighting the limitations of computational power and battery life in wearable devices and the potential benefits of offloading tasks to more capable edge devices. In \cite{rawassizadeh2014wearables} authors have examined the emergence of wearables, and specific smartwatches, and evaluated their stage of adoption and impact on user experiences. In \cite{ometov2021when} authors have delved into the convergence of wearable technology and computing in future networks, exploring the challenges faced by wearables and proposing potential solutions for their development. In \cite{abbas2017mobile} authors have conducted a comprehensive survey on mobile edge computing, emphasizing the role of edge servers co-located with Base Stations and Access Points in offloading computationally intensive tasks from wearables and conserving their limited resources. Authors in \cite{chen2020task} have investigated task offloading for mobile edge computing in software-defined ultra-dense networks, examining the benefits and challenges associated with offloading tasks to edge devices.

In addition, to evaluating the performance of task offloading in general, we delve into a specific use case that revolves around indoor areas such as airports. Security agents at airports often utilize smart glasses to enhance their monitoring capabilities and ensure public safety. These smart glasses enable security personnel to check individuals' body temperatures, potentially detecting those who may have elevated temperatures, a potential symptom of illnesses like COVID-19. By offloading the computing tasks associated with temperature analysis to remote computing resources, the security agents can efficiently process and analyze the data in real-time, enhancing their ability to identify individuals who might pose a risk to public health.
The primary objective of this research paper is to analyze and evaluate different task-offloading scenarios specifically for monitoring applications performed on smart wearable glasses. By offloading resource-intensive tasks to network-edge devices, we aim to improve the performance and usability of these applications. Our study focuses on realistic conditions that are relevant to real-world scenarios, including task completion time, computing capabilities, and energy consumption. By examining these factors, we identify the optimal conditions for effective and efficient task offloading.

Our contributions to this study are summarized as follows

\begin{itemize}

\item Present two-tier edge infrastructure for task offloading, thoroughly assessing its efficacy in enhancing task execution and elevating the overall user experience.

\item Explores the performance limitations of task execution on wearable devices, specifically focusing on the challenges and restrictions encountered during the processing of computationally complex tasks.

\item Investigates the conditions under which task offloading to the nearby computing resources can lead to improvements, examining the potential benefits and drawbacks of offloading tasks from wearable devices to a two-tier edge infrastructure consisting of a mobile device and an edge server.


	

\end{itemize}

The rest of the paper is organized as follows: Section \ref{sec.2}  reviews existing research. Section \ref{sec.3}  describes the system modeling and problem formulation. Section \ref{sec.4} we present the numerical results and performance evaluation of the proposed model. Finally, in Section \ref{sec.5}, we outlined our main conclusions and offered recommendations for further research.

\section{Related work}
\label{sec.2}

The offloading of intensive computing tasks from wearable devices to surrounding computing resources has gained significant attention in recent research. In this section, we present a summary of relevant works that contribute to this specific area.

Satyanarayanan et al. \cite{satyanarayanan2009case} provides a comprehensive survey on Mobile Edge Computing (MEC) and computation offloading. While not exclusively focused on wearable devices, their work explores the potential of offloading tasks from resource-constrained devices to edge networks, highlighting the benefits and trade-offs associated with offloading in various scenarios.
Lai et al. \cite{lai2018wearable} propose a wearable-cloud framework for offloading computationally intensive tasks from wearable devices to the cloud. Their work aims to overcome the limitations of wearable device resources by leveraging the vast computing power of cloud servers. They present infrastructure and task allocation strategies that maximize energy efficiency and reduce latency in offloading scenarios.
Li et al. \cite{li2016context} introduce a context-aware offloading mechanism for wearable devices. Their work considers the dynamic contexts information, such as network conditions and device capabilities, to intelligently determine when and where to offload tasks. Their approach optimizes the offloading decision process and improves the overall performance of wearable devices.
In the context of offloading to nearby devices, Guo et al. \cite{guo2019collaborative} propose a collaborative offloading framework for wearable devices. Their work enables wearable devices to offload tasks to nearby devices in a cooperative manner. They explore the challenges of task partitioning, resource allocation, and communication protocols to ensure efficient and reliable offloading in dynamic environments.
Kang et al. \cite{kang2018adaptive} present a wearable-to-edge offloading system that leverages edge servers deployed in the vicinity of wearable devices. Their work focuses on minimizing latency and energy consumption by offloading tasks to nearby edge servers. They propose an adaptive offloading mechanism that dynamically selects the most suitable edge server based on network conditions and device capabilities.
In addition to edge computing, fog computing has also been explored for offloading tasks from wearables. Ahmed et al. \cite{ahmed2019fog} propose a fog-based offloading framework for wearable devices. Their work introduces a fog layer between wearables and the cloud, enabling task offloading to nearby fog nodes. The framework considers energy efficiency, latency reduction, and scalability in offloading decisions. in \cite{zhang2020adaptive} Zhang et al. present an adaptive task offloading scheme that dynamically selects the most suitable offloading destination (e.g., mobile device or edge server) based on network conditions and device capabilities. The approach aims to reduce the overall latency and energy consumption. Wang et al. \cite{wenxi} survey human-AI social intelligence and discussed the potential of smart glasses in the integration of hybrid Human-AI for Social Computing. Similarly, Cai et al. \cite{CAI2021505} studied human-robot interactions that can be performed through smart devices such as smart glasses. Aung et al. \cite{vesonet,tcoin,blockvanet} proposed task offloading technique for content caching in the context of wireless vehicular networks. In the same vein, Dhelim et al \cite{trust2vec} discussed trust-based task offloading in large-scale IoT systems.  

These studies contribute to the understanding and advancement of offloading tasks from wearable devices to surrounding computing resources. They address various aspects, including edge computing, cloud offloading, context awareness, collaboration, and fog computing. The insights gained from these works provide a foundation for our research on efficient offloading strategies, considering the limitations of wearable device computations and leveraging the available surrounding computing resources.

\section{System Assumptions And  Model Formulation}
\label{sec.3}

\begin{figure}[htb] 
	\centering
	\includegraphics[width=3.0 in]{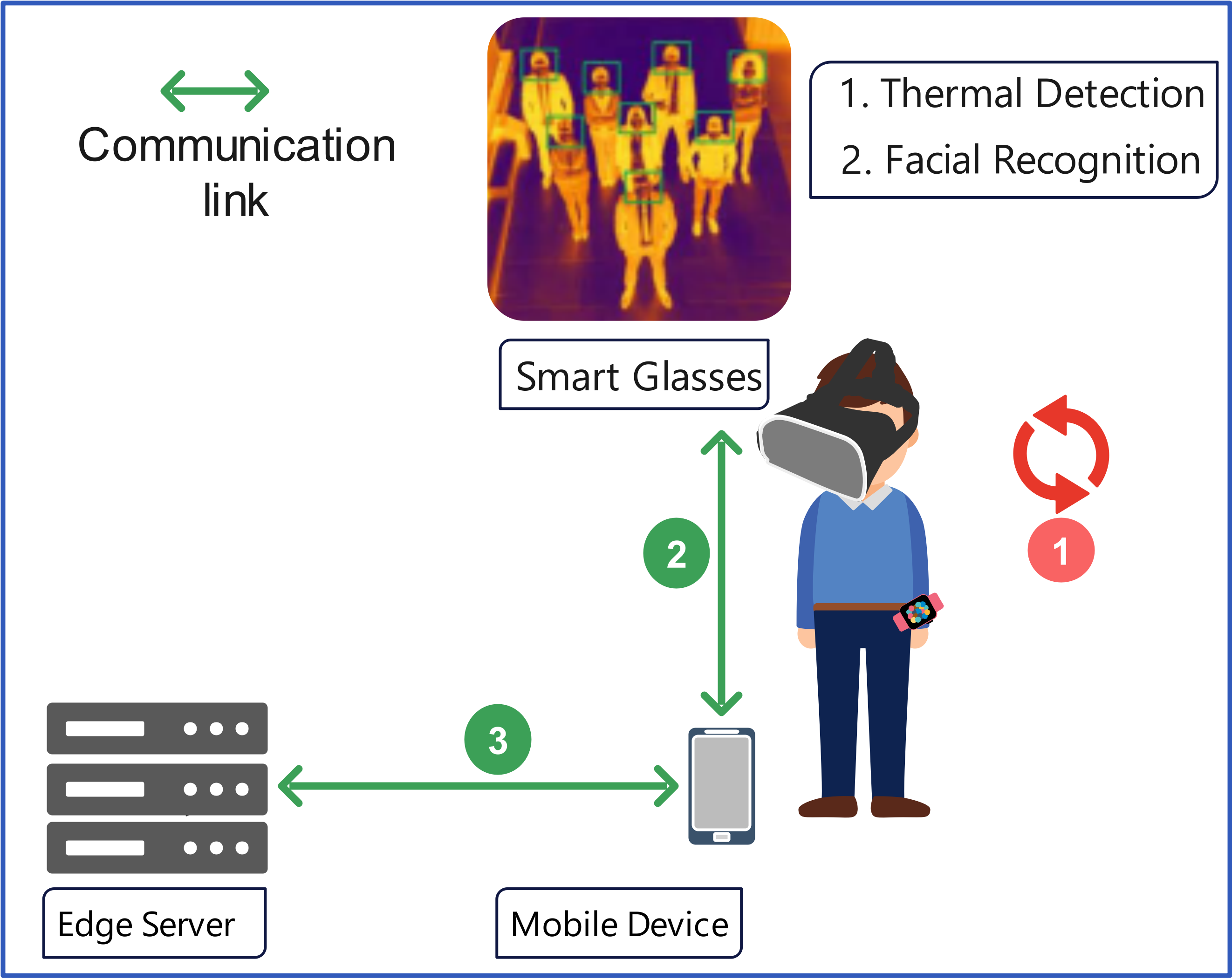}
	\caption{Edge-Mobile Task Offloading Based For Detection of Elevated Body Temperature}
	\label{fig:figure1}
\end{figure}

This section is organized into two distinct parts. The first part of the section will focus on introducing the reference infrastructure being proposed. This will help to understand the underlying structure and design of the system that is being evaluated. In the second part of the section, we will present the key assumptions and mathematical formulas used to calculate the relevant performance metrics for the study. By outlining these aspects, we aim to provide a comprehensive understanding of the methods they used to evaluate the system's performance under investigation.

\subsection{infrastructure Model Description}

In order to provide a clearer illustration of the concept, our study adopts a two-layer infrastructure. The first layer comprises high-performance computing nodes, specifically powerful servers. The second layer encompasses mobile devices and wearable devices, namely smartphones and smart glasses. The smart glasses establish a wireless connection with the user's mobile device, which serves as an interface to the internet. Furthermore, the mobile device establishes a connection with the edge layer, as depicted in Figure \ref{fig:figure1}. This layered architecture facilitates seamless communication and data transfer between the wearable devices and the edge layer, enhancing the overall system functionality. In the context of this study, we specifically focus on the smart glass as the wearable device responsible for executing computationally intensive tasks, such as image and video processing/streaming for monitor applications. This functionality enables users to capture image streams or record videos while on the move.


This study focuses on the characterization of computing tasks based on two key parameters: the captured data size (D in bits) and the computational effort required for bit-level data manipulation using the CPU (C). 

This study specifically focuses on characterizing computing tasks using two essential parameters: the size of the captured data (D in bits) and the computational effort required for bit-level data manipulation using the CPU (C). To accurately assess the computational requirements of the system, authors in \cite{8716520} employed a program profiler. This program profiler effectively tracks and analyzes various program parameters, such as execution time, memory usage, thread CPU time, instruction count, and function calls. By leveraging this information, the study enables an accurate estimation of the number of CPU cycles necessary to process each bit of data for a given task. This comprehensive approach provides valuable insights into the computational demands of the system and facilitates a thorough evaluation of its performance capabilities.


In the context of the illustrated infrastructure depicted in Figure~\ref{fig:figure1}, our study explores three distinct scenarios. In scenario 1, image processing tasks are performed directly on the wearable device, which can be computationally demanding. Given this challenge, we investigate the potential benefits of offloading these tasks to nearby devices, specifically a mobile device (scenario 2) or an edge server (scenario 3). Our primary focus is to assess whether task offloading can effectively conserve energy resources while still satisfying the latency requirements of the application. To evaluate the effectiveness of different offloading scenarios, the study considers three performance metrics: energy consumption, task completion time, and overall user experience. In the context of different scenarios considered, the first scenario stands out with a longer task completion time attributed to the constrained computational capabilities of the wearable device. However, alternative possibilities emerge in the subsequent scenarios, offering increased computation capabilities to overcome this limitation. It is important to note that these enhanced computational capabilities come at the cost of higher energy consumption, primarily associated with the transmission of data between the task operator and the wearable device.

\subsection{System Assumptions And Mathematical Formulations}

Our study is based on several key assumptions. Firstly, Considering the limitations of low-power technologies like Bluetooth/BLE in terms of data rates and communication delays, we focus on Wi-Fi connectivity for efficient task offloading in indoor scenarios. secondly, We assume that the wearable device establishes a Wi-Fi connection with the user's mobile device, which then utilizes a wifi network to access the edge server. thirdly, we specifically consider computationally intensive tasks that cannot be easily divided into subtasks. Lastly, for certain applications such as face detection or automatic license plate recognition, where the results data size is significantly smaller than the entered data. We presume that the transfer time from the computing entities—such as the mobile device and remote server—to the wearable device is negligible. The wearable device incorporates a thermal imaging sensor, which plays a crucial role in analyzing the captured video and extracting the target body image. By identifying specific regions or pixels that represent the human body, the desired image is extracted. Subsequently, this extracted data can be transmitted to a mobile device or an edge server for further processing or analysis, depending on their respective processing capabilities and the selected offloading option. This approach allows for efficient utilization of resources and enables additional tasks or analysis to be performed on the extracted images, enhancing the overall value derived from the data.

We employ mathematical techniques to accurately compute and compare the crucial metrics of interest in three distinct scenarios. These scenarios involve performing the operation directly on the smart glass device, performing it on the user's mobile device, or performing it to an edge server. Specifically, we analyze and quantify the task completion time and energy consumption associated with each scenario. To enhance clarity and facilitate understanding,



For the first scenario, smart glasses execute the task without any offloading. This means that the device operates independently and completes the entire task locally. We can calculate the time it takes to complete the task on the smart glasses, which we refer to, as the task completion time $T_g$, using the following formula:

\begin{equation}
T_g=\frac{D \times C}{F_g}
\end{equation}

Here, $F_g$ represents the computational power of the smart glasses device, expressed in processing cycles per second.

When analyzing the performance of glasses devices during task execution, one important metric to consider is energy consumption. 
The energy consumption of a smart glasses device is accurately defined by the multiplication of the power required by the device during the execution of a task and the duration of that task. This formulation precisely represents the energy utilized by the device during the performance of a specific task. Mathematically, this can be expressed as follows:
\begin{equation}
Eng_g=P_g \times T_g
\end{equation}

Where  $P_g$ refers to the power consumption of the CPU and  $T_g$ task completion time

In the second scenario, the glasses device is typically linked to the user's mobile device, which offers greater resources compared to the glasses resource, as depicted in Figure \ref{fig:figure1}. In this scenario tasks will offload to the mobile device instead of executing locally. Hence, To determine the time required for completing a task on a mobile device, $T_m$, we need to calculate the time required for two components: transferring the captured data from smart glasses to the mobile device over Wi-Fi, $T_{d,m}$, and executing task delay on the mobile device $T_{ex,m}$.

\begin{equation}
T_m=T_{d,m} + T_{ex,m}
\end{equation}

Transferring input data time can be expressed as follow :

\begin{equation}
T_{d,m}=\frac{D }{Bandwidth}
\end{equation}

Executing task delay can be expressed as follow :

\begin{equation}
T_{ex,s}= \frac{D \times C}{F_m}
\end{equation}

To evaluate the total energy usage involved in offloading tasks, it's essential to consider the following scenarios:  1) The consumed energy during transmitting the captured data from the smart glasses device to the mobile device,  2) The consumed energy by the mobile device to acquire the captured data from the smart glasses, 3) the consumed  energy by the mobile device for performing the computing task, and 4) The energy expended by the glasses device during the inactive state when the task is being executed on the mobile device.

\begin{equation}
Eng_m = Eng_{t,g} + Eng_{r,m} + Eng_{ex,m} + Eng_{g_{idle}}
\end{equation}



In the last scenario, computationally demanding tasks are offloaded to an edge server, offering enhanced efficiency compared to local execution on the glasses device. Figure \ref{fig:figure1} illustrates this setup, where the mobile device serves as a gateway node during the offloading process. It receives the captured data from the glasses device and forwards it to the edge server. Furthermore, the mobile device receives the resulting output data from the edge server and relays it back to the smart glasses device.

The completion time of offloaded tasks from the glasses device to the edge server can be referred to as the task computation time and can be expressed as:
\begin{equation}
Time_e = Time_{d,m} + Time_{d,e} +  Time_{ex,e} 
\end{equation}

where $Time_{d,e}$ denotes the time required to offload a task from the mobile device to the edge server, while $Time_{ex,e}$ denotes the time spent executing the task at the edge server. 

To determine the overall energy consumption involved in task offloading to an edge server for execution, the following formula can be employed:

\begin{equation}
Eng_e=Eng_{t, g}+Eng_{r, m}+Eng_{t, e}+Eng_{ex, e} \\ +Eng_{g, idle}+Eng_{m, idle}
\end{equation}

Where $Eng_{t, e}$ accounts for the energy consumed by the mobile device in transmitting the input data to the edge server, $Eng_{m, idle}$ represents the energy used by the mobile device while idle during the execution of the task at the edge server, $Eng_{g, idle}$ refers to the energy used by the glasses device during idle state, and $Eng_{ex, e}$ represents the energy consumed by the edge server in executing the task.

\section{EXPERIMENTAL AND RESULTS}
\label{sec.4}

In this study, we aimed to evaluate the performance of a two-tier edge infrastructure for task offloading from a smart glasses device to the edge in the context of an application that involves capturing people's images on live streaming to detect their temperature. The infrastructure consisted of a smart glasses device, a mobile device, and an edge server. We conducted experiments to measure the task completion time and energy consumption of three different scenarios: executing the task directly on the glasses wearable device, offloading the task to the mobile device, and offloading the task to the edge server. As glasses devices have limited computing resources, we hypothesized that offloading the task to a more powerful device would result in improved task performance. Our experimental results provide insights into the effectiveness of these three scenarios for this specific application and highlight the potential benefits of task offloading for wearable computing in the Internet of Things (IoT) context. simulation parameters are summarized in Table \ref{tab:my-table}.

\begin{table}[]
\centering
\caption{SIMULATION PARAMETERS
}
\label{tab:my-table}
\begin{tabular}{|c|c|lll}
\cline{1-2}
\textbf{Google Glass}  & CPU(0.4GHz to 1.5GHz ) &  &  &  \\ \cline{1-2}
\textbf{Mobile device} & CPU (2.2 GHz )         &  &  &  \\ \cline{1-2}
\textbf{Edge Server}   & CPU (20 GHz )          &  &  &  \\ \cline{1-2}
\textbf{Distance}      & 50 - 600 Meters        &  &  &  \\ \cline{1-2}
\textbf{Data size}     & 0.1 - 2.0 MB           &  &  &  \\ \cline{1-2}
\textbf{Wifi}          & 54Mbps                 &  &  &  \\ \cline{1-2}
\end{tabular}
\end{table}

\begin{figure}[htb] 
	\centering
	\includegraphics[width=3.5 in]{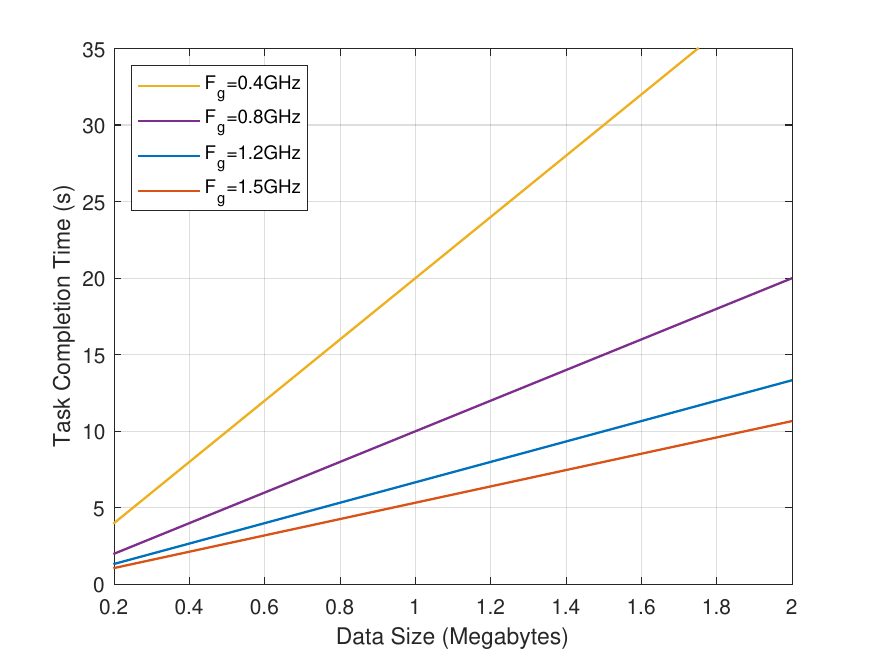}
	\caption{Analyzing task completion time for executing tasks on the glasses device with different CPU frequencies and data entry sizes.}
	\label{fig:figure2}
\end{figure}

\begin{figure}[htb] 
	\centering
	\includegraphics[width=3.5 in]{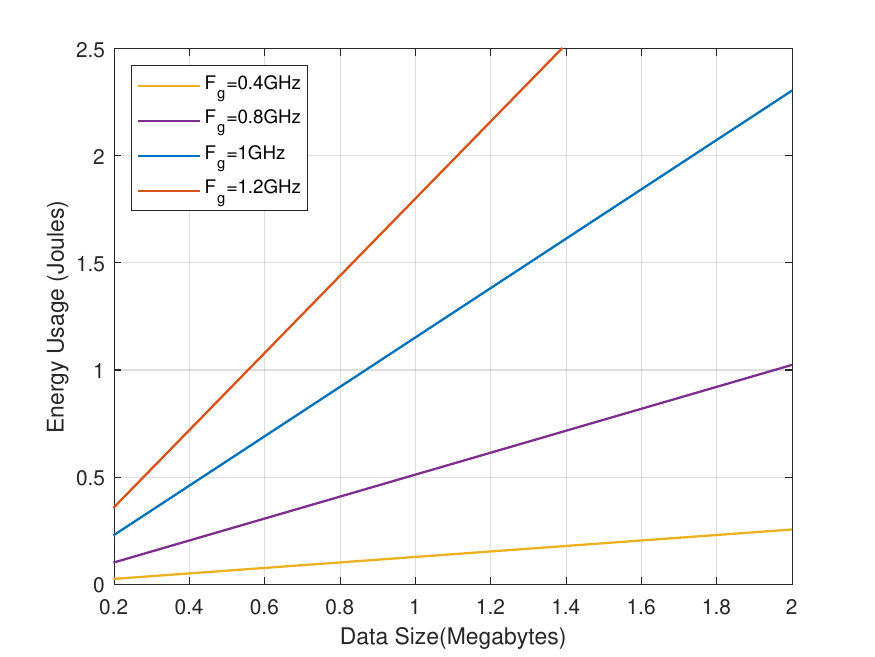}
	\caption{Energy usage for glasses task execution with different CPU rates and data entry sizes.}
	\label{fig:figure3}
\end{figure}

\subsubsection{Smart Glasses: Evaluating Local Task Execution Performance}


  Figure~\ref{fig:figure2} illustrates the task completion time on smart glasses devices without offloading, considering different CPU frequencies ranging from 0.4GHz to 1.5GHz. The figure demonstrates that higher CPU frequencies result in shorter task completion times, indicating that devices with higher computational capacities can process tasks more quickly. Conversely, devices with lower CPU frequencies may experience longer task completion times. This suggests that task offloading can be particularly beneficial for devices with lower computational capabilities, allowing them to offload tasks to more powerful devices and reduce their own processing burden. We observe that the processing time increases linearly with the data size. This linear correlation between the processing time and data size signifies a consistent computational intensity. Regardless of the total data size, the processing time per bit remains unchanged. This insight provides valuable information for predicting and understanding the processing requirements as data sizes scale. Figure~\ref{fig:figure3} illustrates the energy usage on the smart glasses device during local computation. The results indicate that higher CPU frequencies result in lower task execution times. However, As the power consumption is directly inversely correlated to the CPU frequency, it should be emphasized that this performance boost comes at the expense of higher energy consumption.

\subsubsection{Local execution versus edge offloading: a comparative analysis}
Our study examines the total duration of task execution across three different scenarios, each with varying captured data sizes. These scenarios encompass executing the task on the glasses device, offloading the task to the mobile device using a 54Mbps Wi-Fi connection, and offloading the task to an edge server located at distances of 150m, 400m, and 500m from the mobile device via Wi-Fi. Since users can be in different positions within the Wi-Fi coverage area while wearing the device and using the mobile device, the inclusion of different distance settings allows for the anticipated variations in connection performance.

\begin{figure}[htb] 
	\centering
	\includegraphics[width=3.7 in]{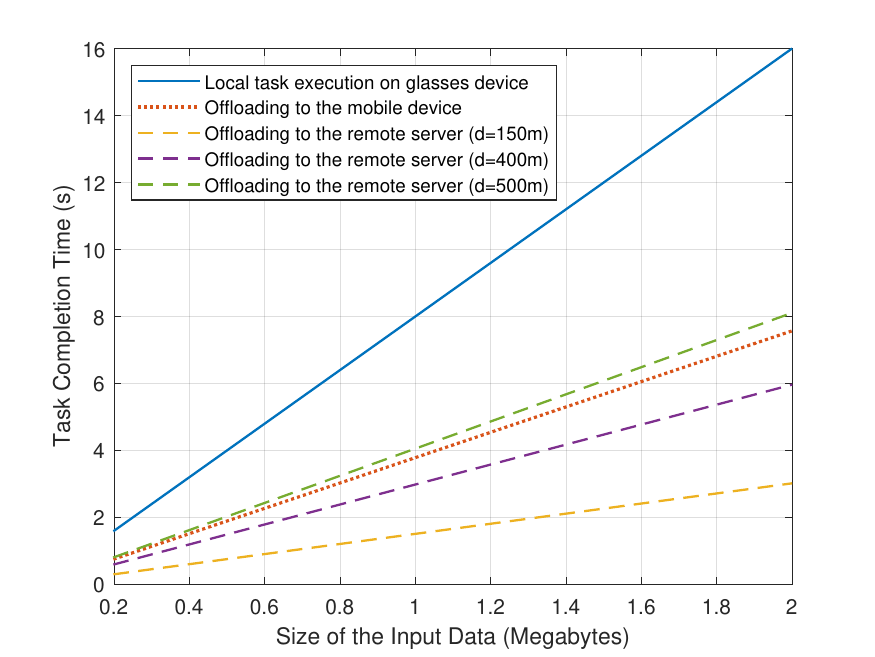}
	\caption{Task Completion Time with Different Captured Data Sizes: Local Execution, Mobile Device Offloading, and Remote Server Offloading (d = 150m, 400m, 500m)}
	\label{fig:figure4}
\end{figure}
\begin{figure}[htb] 
	\centering
	\includegraphics[width=3.7 in]{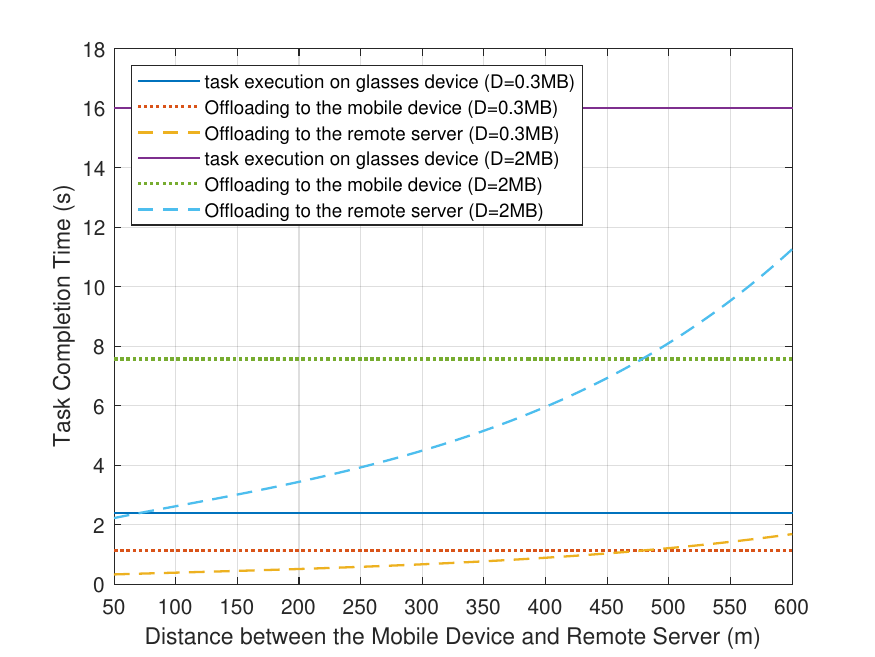}
	\caption{Task Completion Times: Varying Data(0.3MB and 2MB) Sizes and Distances Across Different Scenarios}
	\label{fig:figure5}
\end{figure}
Figure~\ref{fig:figure4} illustrates that task execution time rises with more significant data sizes in all cases. Local task execution on the smart glasses device performs the longest completion time with various data sizes due to limited computational resources.  On the other hand, offloading to the edge server shows the best performance when the user is closer to the wifi access point, benefiting from high data rates and abundant computational resources, and offloading tasks to the mobile device generally show falls in the middle; influenced by different Wi-Fi data rates. In contrast,  offloading to the edge server demonstrates the second-worst performance when the user is far from the access point, mainly due to the degradation of link quality and reduced data rates, which subsequently lead to longer task execution times. Consequently, offloading time-critical tasks can effectively fulfill latency requirements and conserve energy on wearable devices, as long as optimal offloading conditions are provided.

Figure~\ref{fig:figure5} presents the time required to complete task execution by altering the distance between the mobile device and the edge server with different captured data sizes: D=0.30MB for smaller data and D=2MB for larger data. It is evident that offloading to the edge server becomes increasingly expensive as the user moves further away from the remote server, particularly for larger data inputs. This is due to the larger volume of wireless traffic that needs to be exchanged over both short-range and long-range links. Moreover, the results highlight the potential to achieve task completion times below 1s for smaller data when leveraging a nearby edge server for offloading

\subsubsection{Analyzing Task Completion Time and Energy Consumption: Communication vs. Computation}

\begin{figure}[htb] 
	\centering
	\includegraphics[width=3.7 in]{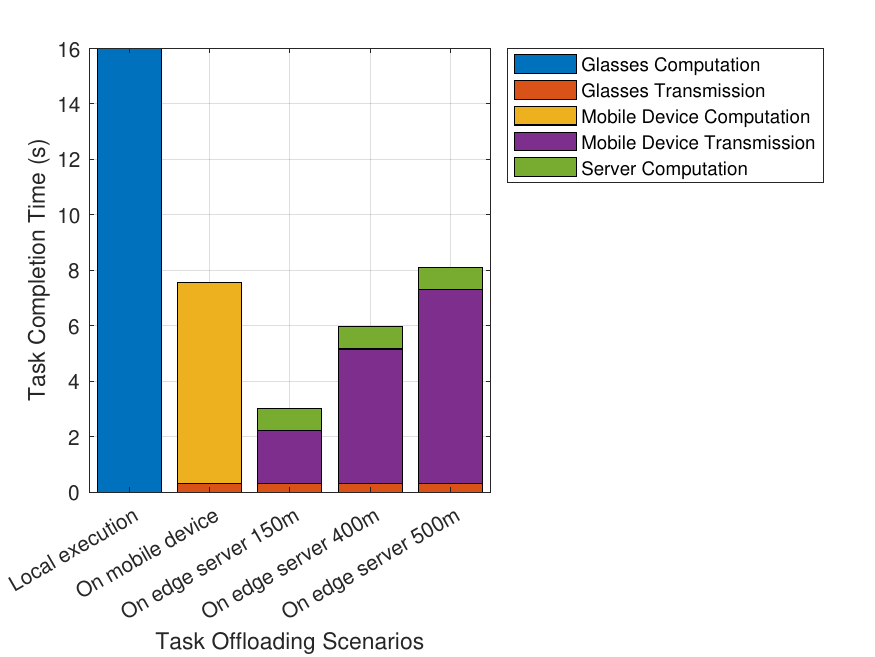}
	\caption{Task Completion Time Comparison for 2MB Data Size: Local Execution, Mobile Device Offloading, and Remote Server Offloading (d = 150m, 400m, 500m)}
	\label{fig:figure6}
\end{figure}

Figure~\ref{fig:figure6} depicts the time required to complete the task under various execution conditions. unexpectedly the time spent on data transfer on the mobile device is longer than on the glasses device. This is due to the access point reduced data transfer rate, which is most noticeable when the user is distant from it. The calculation time at the edge server, on the other hand, is comparatively fast due to its strong resources.

\begin{figure}[htb] 
	\centering
	\includegraphics[width=3.7 in]{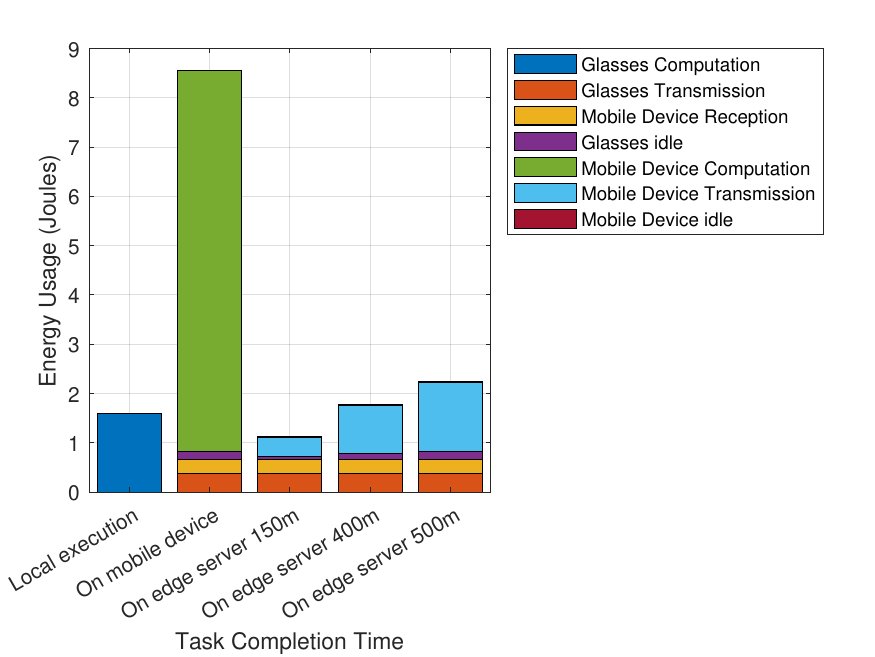}
	\caption{Energy Usage Comparison for 2MB Captured Data Size: Local Execution, Mobile Device Offloading, and Remote Server Offloading (d = 150m, 400m, 500m)
}
	\label{fig:figure7}
\end{figure}

In Figure~\ref{fig:figure7}, the energy usage profile is presented for each of the three task execution scenarios. When the task is executed on the smart glasses device, the energy consumption is solely attributed to the computation performed on the device. However, in the case of assigned tasks  to the mobile device, the overall energy usage comprises several factors. This includes the energy expended by the wearable device during the transfer of captured  data to the mobile device, the energy consumed by the mobile device while receiving the captured data from the glasses device, and the energy consumed by the mobile device during task execution. 

Additionally, it is important to note that while the task is being executed on the mobile device, the wearable device remains in an inactive state, wasting a certain amount of energy.  In an unexpected turn, the computational aspect consumes more energy than communication in the second case. When the task is offloaded to a remote server, the smart glasses device consumes energy to transmit captured data to the mobile device. The mobile device, on the other hand, consumes energy to receive data from the glasses device through a short-range link, and additional energy is used to transmit data to the edge server through a long-range link. In this scenario, the glasses device utilizes some idle energy until the task's output is sent back to it via the mobile device. Meanwhile, when the task is being performed on the remote server, the mobile device remains idle.

\section{Conclusion}\label{c}
\label{sec.5}

This research aims to evaluate the effectiveness of task offloading for body-worn device such as "Smart Glasses" within a two-tier edge infrastructure. In this infrastructure, a mobile device and a remote server are utilized as task processors. Our results indicate that offloading the tasks to the mobile device is consistently more advantageous compared to local execution on the wearable device. This approach not only helps conserve the wearable's limited energy resources but also leads to reduced delays in completing the tasks.  In situations where the mobile device is situated at a considerable distance from network boundaries and encounters unfavorable propagation conditions, it is often more beneficial to carry out the task directly on the mobile device.

This helps to minimize the time required to complete the task unless there are constraints such as a low battery or high computational demands. However, in most scenarios, offloading the task to the remote server is the preferred option compared to executing it on a mobile device. In cases where the tasks are not computationally intensive, it is more advantageous to perform them on the wearable device rather than offloading them. This is due to the significant delay introduced by transferring the wearable device data over wireless networks, which can dominate the overall task completion time. In future research, we plan to conduct experiments-based studies to validate our theoretical analysis and explore opportunities for jointly optimizing task completion time and energy consumption for both wearables and mobile devices. Additionally, investigates the feasibility of a split computing approach, where tasks are partially executed at the remote server and partially on the mobile device or wearable in indoor and outdoor areas with recent network generations.

\section*{Acknowledgment}
This work was supported by The National Natural Science Foundation of China under Grant 61872038.

\bibliographystyle{IEEEtran}
\bibliography{sample}

\vskip 0pt plus -1fil
\begin{IEEEbiography}[{\includegraphics[width=1in,height=1.25in,clip,keepaspectratio]{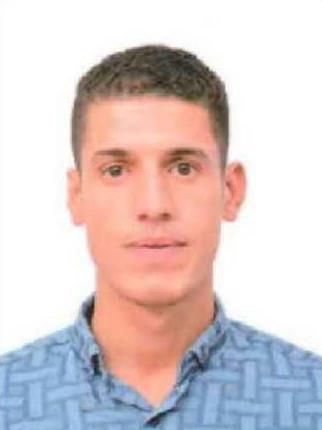}}]{Abdenacer Naouri}
He is currently a Ph.D. candidate at the University of Science and Technology Beijing China, Beijing, China. He 
received his B.S. degree in computer science from the University of Djelfa Algeria, in 2011, and the  M.Sc. degree in networking and distributed systems from the University of Laghouat Algeria, Laghouat, Algeria, in 2016. His current research interests include Cloud computing, Smart communication, machine learning, Internet of vehicles  and Internet of Things. \end{IEEEbiography}

\vskip 0pt plus -1fil

\begin{IEEEbiography}[{\includegraphics[width=1in,height=1.25in,clip,keepaspectratio]{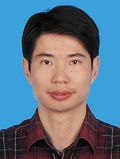}}]{Huansheng Ning} Received his B.S. degree from Anhui University in 1996 and his Ph.D. degree from Beihang University in 2001. Now, he is a professor and vice director of the School of Computer and Communication Engineering, University of Science and Technology Beijing, China. His current research focuses on the Internet of Things and general cyberspace. He is the founder and chair of the Cyberspace and Cybermatics International Science and Technology Cooperation Base. He has presided many research projects including Natural Science Foundation of China, National High Technology Research and Development Program of China (863 Project). He has published more than 150 journal/conference papers, and authored 5 books. He serves as area editor for IEEE Internet of Things Journal (2020-2022), and editor role for some other journals. He is a visiting chair professor of Ulster University, UK.
\end{IEEEbiography}

\vskip 0pt plus -1fil
\begin{IEEEbiography}[{\includegraphics[width=1in,height=1.25in,clip,keepaspectratio]{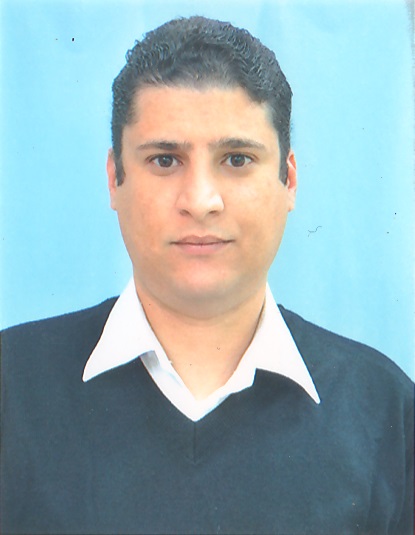}}] {Nabil Abdelkader Nouri}
Is an associate proffessor in University of Djelfa, Algeria. Received his BS's degree in computer sciences from the university of Laghouat, Algeria, in 2003 and his magister degree in networking and distributed systems from the university of Bejaia, Algeria, in 2007. His current research interests include Wireless Networking Design, Internet of Things, Performance Evaluation, Fog Computing and Optimization.
\end{IEEEbiography}

\vskip 0pt plus -1fil
\begin{IEEEbiography}[{\includegraphics[width=1in,height=1.25in,clip,keepaspectratio]{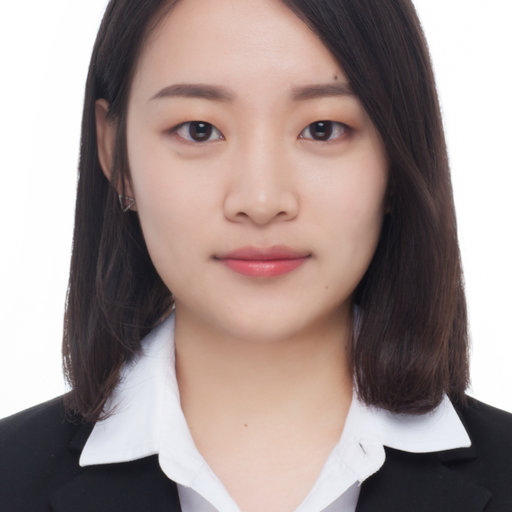}}]
{Feifei Shi} received the B.S. degree from China University of Petroleum, Beijing, China, in 2016, and the M.S. degree from the University of Science and Technology Beijing, Beijing, in 2019, where she is currently pursuing the Ph.D. degree with the School of Computer and Communication Engineering.,Her current research interests include Internet of Things, artificial intelligence, and smart health.
\end{IEEEbiography}

\vskip 0pt plus -1fil
\begin{IEEEbiography}[{\includegraphics[width=1in,height=1.25in,clip,keepaspectratio]{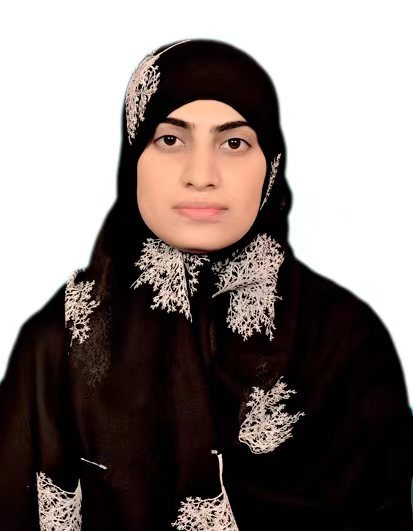}}]{Attia Qammar}(B20200693@xs.ustb.edu.cn) received her BS degree from Bahauddin Zakariya University and her MS from National College of Business Administration and Economics, Pakistan. Currently, she is pursuing her Ph.D. degree from the School of Computer and Communication Engineering at the University of Science and Technology Beijing, China. Her research interests include federated learning, chatbots, data security, and IoT privacy-preserving systems.
\end{IEEEbiography}

\vskip 0pt plus -1fil
\begin{IEEEbiography}[{\includegraphics[width=1in,height=1.25in,clip,keepaspectratio]{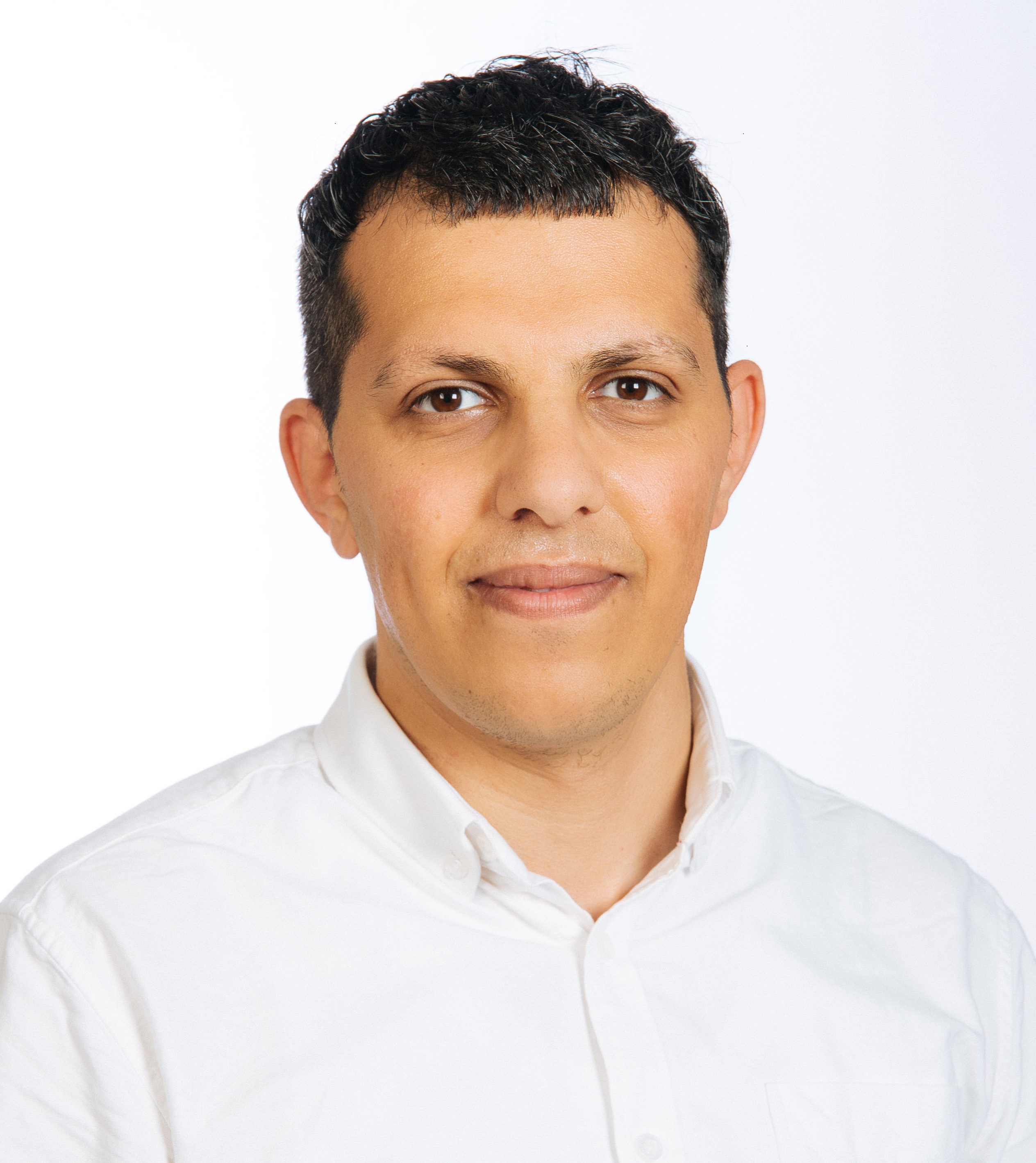}}]{Sahraoui Dhelim} is a senior postdoctoral researcher at University College Dublin, Ireland. He was a visiting researcher at Ulster University, UK (2020-2021). He obtained his PhD degree in Computer Science and Technology from the University of Science and Technology Beijing, China, in 2020. And a Master's degree in Networking and Distributed Systems from the University of Laghouat, Algeria, in 2014. And Bs degree in computer science from the University of Djelfa, in 2012. He serves as a guest editor in several reputable journals, including Electronics journal and Applied Science Journal. His research interests include Social Computing, Smart Agriculture, Deep-learning, Recommendation Systems and Intelligent Transportation Systems.
\end{IEEEbiography}

\vskip -3\baselineskip

\end{document}